\documentstyle[prl,aps,epsfig,floats]{revtex}
\begin{document}
\draft
\newcommand{\beq}{\begin{equation}}
\newcommand{\eeq}{\end{equation}}
\newcommand{\bea}{\begin{eqnarray}}
\newcommand{\eea}{\end{eqnarray}}
\def\lsim{\raise0.3ex\hbox{$\;<$\kern-0.75em\raise-1.1ex\hbox{$\sim\;$}}}
\def\gsim{\raise0.3ex\hbox{$\;>$\kern-0.75em\raise-1.1ex\hbox{$\sim\;$}}}
\def\Frac#1#2{\frac{\displaystyle{#1}}{\displaystyle{#2}}}
\def\al{\alpha}
\def\be{\beta}
\def\ga{\gamma}
\def\de{\delta}
\def\si{\sigma}
\def\C{{\cal{C}}}
\def\O{{\cal{O}}}
\def\wt{\widetilde}
\def\ol{\overline}
\def\l{\left}
\def\r{\right}
\def\no{\nonumber\\}

\setlength{\topmargin}{-.75in} 

\setlength{\evensidemargin}{-0.5in} 
\setlength{\oddsidemargin}{-0.5in}
\setlength{\textwidth}{7in} \setlength{\textheight}{9in}

\twocolumn[\hsize\textwidth\columnwidth\hsize\csname@twocolumnfalse\endcsname

\title{Additional Stringy Sources of the Electric Dipole Moments} 
\author{S. Abel$^1$, S. Khalil$^{1,2}$, and O. Lebedev$^3$}
\address{$^1$~IPPP, University of Durham, South Rd., Durham
DH1 3LE, U.K.\\
$^2$~Ain Shams University, Faculty of Science, Cairo, 11566, Egypt.\\
$^3$~Centre for Theoretical Physics, University of Sussex, Brighton BN1
9QJ, U.K.\\}
\maketitle
\begin{abstract}
We show that string models with low energy supersymmetry
which accommodate the fermion mass hierarchy
generally give non-universal soft trilinear couplings (A-terms).
In conjunction with the apparently large Cabibbo-Kobayashi-Maskawa (CKM) phase,
this results in large fermion EDMs even in the absence of 
CP violating phases in the supersymmetry breaking 
auxiliary fields and the $\mu$-term.
Nonobservation of the EDMs therefore implies that strings select 
special  flavor and/or supersymmetry breaking patterns. 
\end{abstract}
\pacs{PACS numbers: 11.30 Er, 12.60.Jv
\hfill IPPP/01/63 \, \, DCPT/01/126 \, \, SUSX-TH/01-041}
\vskip.5pc
]
\noindent

\vspace{.2in}

Recent measurements of the CP asymmetry
in the $B\rightarrow \Psi K_s$ decay  \cite{Aubert:2001nu} imply that  CP is
significantly violated in nature and  is 
not an approximate symmetry.
Supersymmetric models with large CP phases generally predict 
fermion electric dipole moments (EDMs) that far exceed the experimental limits~\cite{mercury}, and this presents a major challenge for low energy 
supersymmetry, the so called SUSY CP problem.

This problem
arises because there is no {\em a priori} reason for the 
SUSY breaking dynamics to conserve CP.
In particular, in supergravity models the auxiliary fields that break
supersymmetry ($F_S$,$F_T$, etc.) are in general complex, which leads
to complex soft SUSY breaking parameters. These new CP-phases,
which are absent in the non-supersymmetric case, induce large EDMs.
Conventionally, this is 
believed to be the source of the CP problem in this class of
models.  

In this letter, we point out that there is
another source of EDMs inherent in more fundamental 
models such as supersymmetric models derived from strings.
It is present even if the SUSY breaking 
($F_S$,$F_T$, etc.) {\em conserves} CP, and has its origin
in the flavor structures.
String theory, as a fundamental theory, 
has to explain the observed fermion mass hierarchy
and mixings. This generally leaves an imprint on the flavor 
structure of the soft SUSY breaking terms and, when combined 
with the phenomenological 
requirement that the Yukawa matrices contain ${\cal O}(1)$ CP-phases,  
results in unacceptably large EDMs.
The source of the EDMs lies in the CP-phases present already in the non-supersymmetric
case, i.e. those in the Yukawa matrices, which are rendered observable 
by supersymmetry.

Non-observation of EDMs provides a strong constraint
on the fundamental flavor structure of the Yukawas and/or 
supersymmetry breaking, {\em in addition} to constraints on any 
$new$ CP phases that may occur in the latter, and 
indicates that the fundamental theory
selects  quite special patterns of the flavor structures or
supersymmetry breaking.

Let us start with the general arguments that imply non-universality
in the soft trilinear terms relevant to the EDM calculations.
In all supergravity models, the soft SUSY breaking parameters are given in
terms of the K\"ahler potential $K$ and the superpotential $W$. 
In particular, the trilinear parameters are written as~\cite{Brignole:1997dp}
\begin{eqnarray}
&& A_{\alpha \beta \gamma} = F^m \bigl[ \hat K_m  + \partial_m \log Y_{\alpha \beta \gamma}
-\partial_m \log (\tilde K_{\alpha} \tilde K_{\beta} \tilde K_{\gamma} ) \bigr] \;.
\label{A-terms}
\end{eqnarray}
Here the Latin indices refer to the hidden sector fields while the
Greek indices refer to the observable fields; the K\"ahler potential
is expanded in observable fields as $K=\hat K + \tilde K_{\alpha}
\vert C^{\alpha} \vert^2 +...$ and $\hat K_m \equiv
\partial_m \hat K$. The sum in $m$ runs over all of the SUSY breaking fields.
We note that the supergravity notation for the A-terms and Yukawas 
can be connected to the ``usual'' one by fixing the order
of the indices as follows: the first index is to refer to the Higgs fields,
the second to the quark doublets, and the third to the quark singlets, e.g. 
$Y_{{H_1}Q_iD_j}\equiv Y^d_{ij}$.

Since string theory is a fundamental theory, the Yukawa matrices 
must be generated dynamically. That is, 
the Yukawa couplings are functions of fields
that dynamically get vacuum expectation values leading to  the observed
fermion masses and mixings (in the same way that the 
VEV of the dilaton field produces the gauge coupling). 
Also, since CP is a gauge symmetry in strings \cite{Dine:1992ya}
and can be broken only spontaneously,
the fields responsible for generating the Yukawa couplings
must attain CP-violating VEVs in order to produce the CKM phase.
Therefore, the derivatives of the Yukawa couplings in Eq.\ref{A-terms}
will generally be non-zero and should be taken into account.
As we will see, this contribution is  often significant even
if the corresponding SUSY breaking component $F^m$ is small.

To estimate how large these contributions are expected to be,
we will  consider two possible ways to generate the Yukawa hierarchies:
through exponential factors, as in the heterotic string, and 
through non-renormalizable operators, as for example in
type I or other models. 

In weakly coupled heterotic orbifolds,
the  Yukawa couplings are
calculable and given  in terms of the T-moduli fields. 
The fermion mass hierarchy has a geometrical origin and appears due to the fact
that interaction of the states placed at different orbifold fixed points
is exponentially suppressed  \cite{Hamidi:1986vh}. 
The matter fields must
belong to the twisted sectors since otherwise the couplings are 1 or 0 and 
the mass hierarchy cannot be
generated at the renormalizable level.
The Yukawa coupling of the states at the fixed points $f_{1,2,3}$ belonging
to the twisted sectors $\theta_{1,2,3}$ is given by \cite{Casas:1993ac}
\begin{eqnarray}
&& Y_{f_1 f_2 f_3}=
N  \sum_{ {u} \in Z^n}
{\rm exp} \biggl[ -4\pi T \left(
{f_{23}} + {u} \right)^T
M \left( {f_{23}} + {u} \right)
\biggr]\;.
\label{yukawa}
\end{eqnarray}
where $f_{23}\equiv f_2 - f_3$, $N$ is a normalization factor, and the matrix $M$ 
(with fractional entries)
is related to the internal metric of the orbifold.
Here $T$ is normalized such that $T\rightarrow T+i$ is a symmetry of the 
Lagrangian \cite{Lebedev:2001qg}.
For a realistic case, Re$T={\cal O}(1)$, the sum is dominated by one term and
\begin{equation}
\partial_T \ln Y_{ f_1 f_2 f_3 } \simeq -4\pi~
{f_{23}}^{ T} M ~{f_{23} }
\label{lnY}
\end{equation}
 for some (typically fractional) ${f_{23}}$ depending on the positions 
of the fixed points. This expession is independent of $T$ and 
can be evaluated for various orbifolds. 
It vanishes if the fixed points coincide, i.e. there is no suppression
of the Yukawa interaction.
In other cases it is  of order one (or larger, see e.g. \cite{Khalil:2001dr}). 
This creates a significant non-universality
in the A-terms unless $F_T \simeq 0$. 
Indeed, to produce a fermion mass hierarchy we are bound to place
different quark generations at different orbifold fixed points. 
Thus, the expression (\ref{lnY}) will be generation dependent. 
The other contributing terms in Eq.\ref{A-terms},  
$\hat K_T=-3/(T+\bar T)$ and 
$\partial_T \log (\tilde K_{\alpha} \tilde K_{\beta} \tilde K_{\gamma} )
=(n_{\alpha}+n_{\beta}+n_{\gamma})/(T+\bar T)$ are generation independent
since the sum of the modular weights $n_i$ is fixed by the point group selection
rule for the Yukawa interactions.  Unless we encounter a special case $F_T \simeq 0$,
the non-universality will be present.
We thus conclude that the A-term non-universality is a generic feature
of realistic heterotic models. 

In general, 
it might be too strict a requirement to demand that all of the required features appear
at the renormalizable level.
In many string models the renormalizable Yukawa couplings are either unknown
or are 0,1 (e.g. Type I models). Thus, to reproduce the observed
fermion masses, non-renormalizable operators must be taken into account. 
The mass hierarchy is then created via powers of a small (in Planck units) 
VEV of a certain field $\phi$ \cite{Faraggi:1994su},\cite{Abel:2001ur}, i.e.
\begin{equation} 
Y_{\alpha\beta\gamma} \sim \phi^{q_{\alpha\beta\gamma}}\;.
\end{equation}
Here $q_{\alpha\beta\gamma}$ 
is integer and can sometimes be associated with a $U(1)$ charge
(as in the Froggatt-Nielsen mechanism).
Clearly, if this field breaks supersymmetry, $F_{\phi}\not=0$, the generated A-terms
will be nonuniversal:
\begin{equation}
\partial_\phi \ln Y_{\alpha\beta\gamma} = {q_{\alpha\beta\gamma}\over \phi}\;.
\end{equation}
Generally, $\phi$ is expected to give a (small) contribution to supersymmetry breaking.
To estimate its natural size, let us recall that the K\"{a}hler potential for untwisted
fields is given by $K=-\ln(S+\bar S)-3 \ln(T+\bar T- \phi \bar \phi)$.
The resulting auxiliary field is then
\begin{equation}
F_{\phi}=m_{3/2}{T+\bar T-\phi \bar \phi \over 3} \left[ \bar\phi
{W_T\over W} + {W_\phi\over W}  \right]\;.
\end{equation}
A similar result holds for a twisted $\phi$.
For small $\phi$,  $F_{\phi}$ is typically of order $\phi m_{3/2}$ 
(see also e.g.  \cite{Abel:2001ur}).
As a result, the Yukawa-induced contribution to the A-terms is
\begin{equation}
\Delta A_{\alpha \beta \gamma} \sim q_{\alpha\beta\gamma}~ m_{3/2}\;.
\end{equation}
The ``charges''  $q_{\alpha\beta\gamma}$ are generation-dependent
and order one, so the resulting non-universality is very significant.
Again, we come to the conclusion that realistic models of the fermion masses 
entail  non-universal A-terms. We note that additional nonuniversal contributions may come from the K\"{a}hler potential. However, these are not direcly related to the Yukawa structures, so we do not discuss them here. 

On general grounds, it is natural to expect that once non-universality is created 
in the Yukawa interactions,
it will also  appear in the analogous soft-breaking terms. 
This leads to large EDMs even if the SUSY breaking dynamics do not generate new CP phases,
i.e. $F_S$,$F_T$, etc. and the $\mu$-term are real.
To see this let us suppose that initially there are no CP-violating phases in any of the soft
breaking parameters. 
At first sight this seems to avoid
overproduction of the EDMs. However
this is not the case. Indeed, the relevant trilinear interactions 
are 
\begin{eqnarray}
&& \Delta {\cal L}_{\rm soft}= -{1\over 6 } 
A_{\alpha \beta \gamma}  Y_{\alpha \beta \gamma}
C^\alpha C^\beta C^\gamma\;. 
\end{eqnarray}
Denoting  $\hat A_{\alpha \beta \gamma} \equiv 
A_{\alpha \beta \gamma}  Y_{\alpha \beta \gamma}$ and identifying
$\hat A_{{H_1}Q_iD_j}\equiv \hat A^d_{ij}$, it is easy to see that generally
$\hat A$ and Y matrices are ``misaligned''.
In the basis where the Yukawas are diagonal and real (the super-CKM basis), 
the $\hat A$ matrices are 
not diagonal while their diagonal elements contain complex phases. 
Under the superfield rotation 
$\hat U_{L,R} \rightarrow V^u_{L,R}~ \hat U_{L,R} \;,\;
 \hat D_{L,R} \rightarrow V^d_{L,R}~ \hat D_{L,R}$
to the super-CKM basis, i.e. 
$Y^u \rightarrow V_L^{u\dagger}~ Y^u~ V_R^{u}={\rm diag}(h_u,h_c,h_t)$,
$Y^d \rightarrow V_L^{d\dagger}~ Y^d~ V_R^{d}={\rm diag}(h_d,h_s,h_b)$, 
the $\hat A$ matrices transform as
\begin{eqnarray}
\hat A^u \rightarrow V_L^{u\dagger}~ \hat A^u~ V_R^{u}\;,\;
\hat A^d \rightarrow V_L^{d\dagger}~ \hat A^d~ V_R^{d}\;.
\end{eqnarray}
Since the Yukawa matrices  contain ${\cal O}(1)$ CP-phases,
so do the rotation matrices $V_{L,R}$.
Therefore, the rotation by itself  can induce diagonal CP-phases even
if the $A$- or $\hat A$- matrices  initially  are  completely real.  
Furthermore, due to non-universality the diagonal elements of $\hat A$ will 
have contributions from all three 
generations, e.g.
\begin{equation}
\hat A_{11}^u \propto m_u + \epsilon m_c + \epsilon' m_t \;.
\label{a11}
\end{equation}
This is to be contrasted with the universal case when the Yukawas
and $\hat A$ matrices are diagonalized simultaneously and  
 $\hat A_{ii}$ are proportional to $m_i$. 
The imaginary parts of the 
diagonal entries of the $\hat A$ matrix in the super-CKM basis induce quark EDMs
via the gluino-squark loops with LR mass insertions.
Eq.\ref{a11} implies that the magnitude of the mass insertions contributing to the 
EDMs, $(\delta^{u,d}_{LR})_{ii}\sim \hat A^{u,d}_{ii} \langle H_{u,d} \rangle
/\tilde m^2$, is significantly enhanced and even a small CP-phase
can result in large EDMs. 

\begin{figure}[t]
\epsfig{file=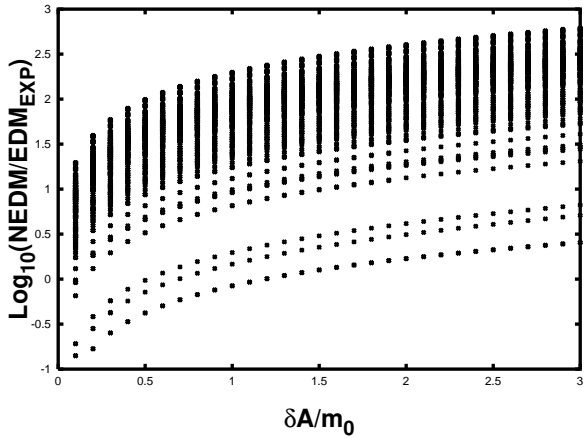, width=8cm}\\

\caption{Neutron EDM versus $\delta A$.}
\label{fig1}
\end{figure}
\begin{figure}[t]
\epsfig{file=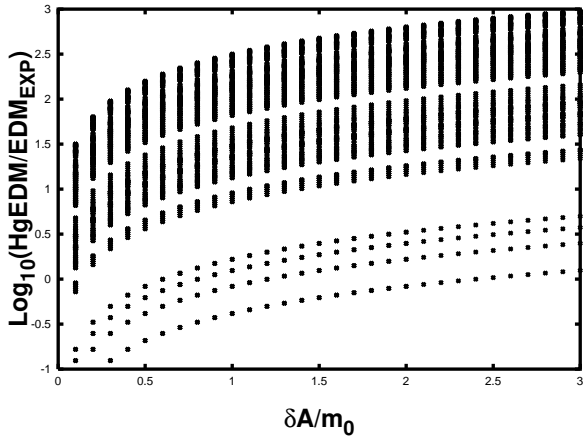,width=8cm}\\

\caption{Mercury EDM versus $\delta A$.}
\label{fig2}
\end{figure}

  
To illustrate the strength of this effect, we present the scatter plots
(Fig.\ref{fig1},\ref{fig2})
of the neutron and mercury EDMs versus $real$ A-terms.
We take representative Yukawa
textures (similar to those of 
Ref.\cite{Abel:2000hn} but possessing no symmetry) 
with $conservative$ flavor mixing, i.e.  
$(V_{L,R})_{13}\sim V_{td}^{CKM}\sim {\cal O}(10^{-2})$,
 $(V_{L,R})_{12}\sim \sin\theta_C$, and containing order one complex phases.
We fix the universal 
scalar mass $m_0$ and $m_{1/2}$ to be 200 GeV 
(i.e. $m_{\tilde q}\sim m_{\tilde g}\sim$ 500 GeV at $M_Z$) and $\tan\beta=3$.
We write $A^{u,d}_{ij}=m_0+\delta A~ x_{ij}$
and vary $\delta A$ from 0 to 3$m_0$
with   $x_{ij}$ randomly selected in  the range $[-1, 1]$. 
To calculate the mercury EDM we use the results of Ref.\cite{Falk:1999tm}
and the bounds on the mass insertions of Ref.\cite{Abel:2001vy}.
The neutron EDM requires Im$(\delta^{u,d}_{LR})_{11} \leq {\cal O}(10^{-6})$,
whereas the bound from the mercury EDM is 
Im$(\delta^{u,d}_{LR})_{11} \leq {\cal O}(10^{-7})$.

We see that the EDMs exceed the experimental bounds by up to 3 orders of 
magnitude. In order sufficiently to suppress this effect, the sfermion 
masses must be pushed up above 10 TeV. This would however create a large 
hierarchy between the electroweak and the SUSY scales leading to a large 
fine-tuning \cite{Abel:2001vy}, so one of the primary motivations for 
supersymmetry would be lost. 

The effect described above seems quite close in spirit to the notorious
FCNC problem. Yet, it is different from the latter since even if the FCNC
are suppressed sufficiently, the EDM problem will still persist. Indeed, the
strongest FCNC constraints come from the kaon physics and, for instance,
$\epsilon_K$ requires 
  $\sqrt{\vert {\rm Im}(\delta^{d}_{LR})_{12}^2 \vert}\leq
3.5 \times 10^{-4}$ for the above parameters \cite{Gabbiani:1996hi} . 
This bound is satisfied automatically in most SUGRA models
with ${\cal O}(1)$ non-universality in the A-terms because these mass insertions are 
suppressed by the quark masses. In contrast, the level of sensitivity of the EDMs
is ${\cal O}(10^{-7})$ and even a 10\% nonuniversality can overproduce EDMs
by an order of magnitude (Figs.1,2).
 
To be specific, let us consider an example. In the $Z_6$-I orbifold model of 
Refs.\cite{Lebedev:2001qg,Khalil:2001dr}, the Higgs fields belong
to the $\theta$ twisted sector, quark doublets -- to the $\theta^2$ sector,
and quark singlets -- to the $\theta^3$ sector (for the fixed point assignment, see 
\cite{Lebedev:2001qg}).  The squark masses are generation--independent since different 
generations belong to the same twisted sector.
The resulting A-terms depend
on the Goldstino angle $\Theta$ which measures the balance between the 
dilaton and moduli contributions to SUSY breaking. For $\vert T \vert \sim 1$ and 
a conservative value
$\tan\Theta \sim -2$ which corresponds to  mostly dilaton SUSY breaking
($F_S/F_{T_i} \sim 10$), the arising A-term texture is
\begin{eqnarray}
A^d \sim -\sqrt{3} m_{3/2} \left( \matrix{ 0 & 0 & -2 \cr
                                           1 & 1& 3  \cr
                                           -1 & 2 & 1
}\right) \;.
\end{eqnarray}
Here $m_{3/2}$ is the gravitino mass and  we have assumed real $F_S$ and $F_T$. The resulting mass insertions are  $(\delta^{d}_{LR})_{11}\simeq -3\times 10^{-5}-3\times 10^{-5}$i,
$(\delta^{d}_{LR})_{12}\simeq -2\times 10^{-5}+2\times 10^{-5}$i,
$\vert (\delta^{d}_{LR})_{23}\vert \sim {\cal O}(10^{-4})$,
$\vert (\delta^{d}_{LR})_{13}\vert \sim {\cal O}(10^{-5})$. 
Thus, the SUSY contribution to $K- \bar K$ mixing is negligible (only
the contribution to $\epsilon'$ is significant).
The FCNC bounds involving the third generation are much looser and easily satisfied.
On the other hand, the neutron and mercury EDMs are overproduced by one-two orders of magnitude.    
We found that these results are qualitatively stable under variation of the 
Goldstino angle. Clearly, this situation is quite general and holds in most models with ${\cal O}(1)$
non-universality in the A-terms.

The additional EDM contributions emerge due to the fact that string theory, as a fundamental 
theory, must explain the fermion mass patterns as well as accommodate a 
large CKM phase. The underlying mechanism generally affects the soft terms 
thereby inducing large EDMs. The two sectors, the Standard Model and the 
supersymmetric sector, which can be  assumed to be  disconnected in the 
context of the MSSM, are tightly related in string models and it is the 
absence of CP violation in one of them and its  presence  in the other 
that leads to the conflict.

In this context, EDMs serve as a probe of supersymmetric flavor physics
as well as supersymmetry breaking.
Non-observation of the EDMs implies that string theory selects 
special flavor structures or a special pattern of SUSY breaking.
The non-universal contributions to the EDMs are absent if 
the A-terms happen to be $universal$. This requires that (1) the 
fields that generate the Yukawa matrices do not break SUSY,
\begin{equation}
F_T \simeq 0 \;\;,\;\; F_\phi \simeq 0\;,
\end{equation}
and (2) that there are no generation-dependent contributions from the 
K\"{a}hler potential in Eq.\ref{A-terms}. This occurs if different moduli
are responsible for SUSY breaking and generating the Yukawa couplings,
yet it remains to be understood why the latter moduli do not break supersymmetry. 
Exact universality can be a result
of dilaton-dominated SUSY breaking. This scenario is known to lead to
charge and color breaking minima which, however, can be avoided by lowering
the string scale to intermediate values.
On the other hand, dilaton domination would be hard to implement in semi-realistic
string models
which exhibit CP violation and dilaton stabilization \cite{Khalil:2001dr},\cite{Abel:2000tf}.

The  problem can also be avoided if the flavor structures are
$hermitian$,
\begin{equation}
Y^a=Y^{a~\dagger} \;\;,\;\; A^a=A^{a~\dagger}\;.
\end{equation}
In this case, the $\hat A$ matrices are also hermitian and,
since $V_L^a=V_R^a$, the hermiticity is preserved by the basis rotation \cite{Abel:2000hn}.
This eliminates the flavor-diagonal phases in the A-terms.
The non-hermitian corrections appear only due to small renormalization group 
effects and lead to naturally suppressed EDMs.
To justify the hermiticity requires 
the presence of  an additional symmetry \cite{Mohapatra:1997su}
and it is a  non-trivial task 
to embed such models in string theory.

The problem becomes milder (yet it exists) if
one assumes that the 
up Yukawa matrix is diagonal while the down one is proportional to the CKM matrix.
In this case, the CP-phase appears only in the (13) and (31) entries of
the down Yukawa matrix. 
This suppresses the effect of the super-CKM rotation,
 yet the mercury EDM often exceeds the limit by about an order of magnitude. 
Another interesting possibility is to have $matrix$-factorizable A-terms:
$\hat A=A \cdot Y$ or $Y \cdot A$. This suppresses the magnitude of the mass insertions
since now there is a contribution from only one generation, $(\delta_{LR})_{ii} \propto m_i$.
However, again it is not clear whether these special cases can be obtained in 
realistic string models.
In any case, the above considerations provide a  selection rule: 
any realistic string model must satisfy the requirement of absence 
of the non-universal contributions to the EDMs.

To summarize, we have argued that string models which accommodate
the fermion mass hierarchy and mixings generally lead to large
EDMs even in the absence of CP-phases in the SUSY breaking auxiliary
fields and the $\mu$-term.
Non-observation of the EDMs implies that the supersymmetric 
structures have a special flavor or SUSY breaking pattern,
which can be probed in  current and future particle experiments.   
We have benefited from valuable discussions with G. Kane.

\end{document}